\documentclass[letter]{aa}     % for the letters
\usepackage{graphicx}
\usepackage{txfonts}
\usepackage{lipsum}
\usepackage{subcaption}         % necessary for continued figures, example in section 3
\usepackage{lscape}             % to rotate a single page table, example in appendix.
\usepackage{placeins}           % useful with \FloatBarrier, to keep 
\usepackage{hyperref}
   \usepackage{caption} 

\begin{document}

%%%%%%%%%%%%%%%%%%%%%%%%%%%%%%%%%%%%%%%%
% if you use custom commands in your title,
% ensure to check your title when submitting!
%%%%%%%%%%%%%%%%%%%%%%%%%%%%%%%%%%%%%%%%
   \title{Discovery of thermal radio recombination line emission towards the high-mass protostar IRAS18162$-$2048}

   % \subtitle{Subtitle}

%%%%%%%%%%%%%%%%%%%%%%%%%%%%%%%%%%%%%%%%
% Please separate each author with the \and command
%
% Use the \corrauth to provide the corresponding
% author address. It will be automatically inserted as 
% footnote in the PDF output.
%
% Please DO NOT include ORCIDs next to author names.
% Instead, please provide an active address for each coauthor:
% it will be automatically extracted by EDPS editorial system, 
% and co-authors will be be able to authenticate their ORCID.
%
% Only authenticated ORCIDs will be taken into account.
% ORCIDs included here will be removed.
%%%%%%%%%%%%%%%%%%%%%%%%%%%%%%%%%%%%%%%%

   \author{R. Fedriani\inst{1}\corrauth{fedriani@iaa.es}
        \and J.F. G\'omez\inst{1}
        \and Y. Cheng\inst{2}
        \and M. Osorio\inst{1}
        \and G. Anglada\inst{1}
        \and C-Y. Law\inst{3}
        \and J.M. Masqu\'e\inst{4}
        \and A.F. Placinta-Mitrea\inst{1}
        \and G. Bl\'azquez-Calero\inst{1}
        \and G. Cosentino\inst{5}
        }

   \institute{Instituto de Astrof\'isica de Andaluc\'ia, CSIC, Glorieta de la Astronom\'ia s/n, E-18008 Granada, Spain\\
              \email{fedriani@iaa.es}
            \and
            National Astronomical Observatory of Japan, 2-21-1 Osawa, Mitaka, Tokyo 181-8588, Japan
            \and
            Osservatorio Astrofisico di Arcetri, Largo Enrico Fermi, 5, 50125 Firenze FI, Italy
            \and
            Departament de F\'isica Qu\`antica i Astrof\'isica, Institut de Ci\`encies del Cosmos (ICC), Universitat de Barcelona (IEEC-UB), Mart\'i i Franqu\`es 1, 08028 Barcelona, Spain
            \and
            Institut de Radioastronomie Millim\'etrique, 300 Rue de la Piscine, 38400 Saint-Martin-d`H\`eres, France
    }

   \date{Received ---; accepted ---}

% \abstract{}{}{}{}{}
% 5 {} token are mandatory
 
  \abstract
   {The physical conditions within the innermost few tens of astronomical units of high-mass protostars ($M_*>8~M_\odot$) remain largely unexplored. At these scales, ionised emission can trace photoionised gas, accretion disks, or the base of protostellar jets.}
   {We search for ionised emission in the form of radio recombination lines (RRLs) in high-mass star-forming regions.} 
   {We used archival data from the Atacama Millimeter/submillimeter Array (ALMA) interferometer in band 6 ($\sim$1.3~mm) with a resolution of $\theta_\mathrm{beam}\sim53~\mathrm{mas}\times42~\mathrm{mas}$ to examine the high-mass protostar IRAS18162$-$2048, the driving source of the Herbig-Haro (HH) objects HH80-81.}
   {We detect, for the first time in this protostar, the thermal emission of the RRL H30$\alpha$ ($\mathrm{rest~frequency}=231.90093$~GHz). The ionised emission is spatially unresolved in our observations, but we estimate an emitting region of $\lesssim50$~au at the source distance of 1.4~kpc. The H30$\alpha$ line profile is broad and spectrally resolved with a local standard of rest velocity of $\sim-14~\mathrm{km~s^{-1}}$ (i.e. $\sim-26~\mathrm{km~s^{-1}}$ blueshifted with respect to the systemic velocity of the cloud) and a full width at half maximum of $\sim82~\mathrm{km~s^{-1}}$. These observations suggest that the H30$\alpha$ line is tracing outflowing blueshifted material in the well-known radio jet from this object. This detection opens the possibility of detailed studies of the kinematics and physical properties of ionised jets.}
   {}

   \keywords{Stars: formation --
             Stars: individual: IRAS 18162-2048 --
             ISM: individual objects: HH 80-81 --
             ISM: jets and outflows
               }

   \maketitle
   \nolinenumbers

%%%%%%%%%%%%%%%%%%%%%%%%%%%%%%%%%%%%%%%%%%%%%%%%%%%%%%%%%%%%%%
\section{Introduction}

    Growing evidence suggests that the formation of high-mass stars shares many similarities with that of their low-mass counterparts \citep[e.g.][]{beuther2025}. A key difference, however, is that massive protostars can photoionise their immediate surroundings, while additional ionisation may arise from shocks in accretion flows and protostellar jets \citep{bally2016,anglada2018}.

    The high extinction towards massive protostars makes radio observations a unique probe of their innermost ionised environments. In particular, radio recombination lines (RRLs) offer direct access to the kinematics of ionised gas \citep{dupree1970,brown1978}. While RRLs are routinely detected towards \ion{H}{ii} regions \citep{reifenstein1970,wilson1970}, detections towards protostars remain comparatively rare. Most of the detected RRL emission in protostars is of a maser nature \citep[see e.g.][]{jimenez-serra2011,zhang2019nature,zhang2019G45,zhang2022,law2022,martinez-henares2024}. However, \cite{sanchez-monge2025} presented the first possible example of thermal RRL emission associated with a radio jet, in the W75N(B)-VLA3 protostar.

    The high-mass star-forming region GGD27 \citep{gyulbudaghian1978} is located at $d\sim1.4$~kpc and has $\sim10^4~L_\odot$ \citep{gomez2003,anez2020}. Its main source, IRAS18162$-$2048 (also known as GGD27-MM1), is a $\sim20~M_\odot$ B0 protostar \citep{anez2020} that powers the parsec-scale \citep{marti1993,marti1995,masque2012,mohan2023obs} HH80-81 outflow \citep{reipurth1988}, the fastest protostellar jet known, with velocities exceeding $1000~\mathrm{km~s^{-1}}$ \citep{marti1993,marti1995,bally2023}. The first centimetre continuum emission associated with the region was detected by \citet{rodriguez1980}, while subsequent higher-angular-resolution observations revealed an elongated radio source towards IRAS18162$-$2048 and the first radio counterparts of the HH80-81 objects \citep{rodriguez1989}. Millimetre continuum observations later uncovered compact dusty structures surrounding the protostar \citep{gomez2003,qiu2009}, and subsequent studies established the presence of a dusty accretion disk and a rotating molecular structure at the core of the system \citep{fernandez-lopez2011dust,fernandez-lopez2011molecule,carrasco-gonzalez2012,girart2017,girart2018,anez2020}. More recently, \citet{fedriani2026} detected the first hydrogen recombination line in the GGD27 region through Br$\gamma$ emission towards IRS~7. However, IRAS18162$-$2048 remained undetected at wavelengths shorter than 2.45~$\mu$m, and no RRL has yet been reported towards this source.

    In this Letter, we report the first detection of RRL emission towards the high-mass protostar IRAS18162$-$2048 through the H30$\alpha$ transition. We detect compact, blueshifted ionised gas and present evidence that the emission is associated with the innermost regions of the HH80-81 jet.    

%%%%%%%%%%%%%%%%%%%%%%%%%%%%%%%%%%%%%%%%%%%%%%%%%%%%%%%%%%%%%%
\section{Observations and data reduction}\label{sect:observations}

    Observations at $\sim1.3$~mm (band 6) were obtained with the Atacama Large Millimeter/Submillimeter Array (ALMA) on 2017 September 11 and 2019 July 28 (Project ID: 2016.1.01036.S, PI: P. Sanhueza). Data from both epochs were combined to obtain the final dataset. The range of antenna baseline lengths were 41.4 to 7552.3~m and 92.1 to 8547.9~m in 2017 and 2019, respectively. Four spectral windows (SPWs) were selected in both epochs, centred at 217.8, 220.0, 232.0, and 234.5 GHz, each with a bandwidth of 1.875 GHz. They were sampled with channel widths of 488.3 kHz (SPWs at 217.8 and  220.0 GHz) and 979.6 kHz (SPWs at 232.0 and 234.5 GHz). The phase centre of the observations was set at R.A.(ICRS) $= \mathrm{18^h19^m12.1983^s}$ and Dec.(ICRS) $=-20^\circ47'29.020''$. Calibrated data were provided by the observatory. Further data reduction and imaging were conducted with the Common Astronomy Software Applications package \citep[CASA,][]{mcmullin2007}, version 6.5.4-9. 

    We selected all line-free channels of the four SPWs to construct a continuum dataset. This continuum was self-calibrated in phase and amplitude and the resulting corrections were applied to the whole dataset (including line emission). The final continuum image was obtained with the CASA task \texttt{tclean}, in multi-term, multi-frequency synthesis mode and applying Briggs weighting \citep{briggs1995} with a robust parameter of 0.5. In this image, centred at 226.203 GHz, the synthesised beam has a full width at half maximum (FWHM) of $\theta_s=53~\mathrm{mas}\times42~\mathrm{mas}$, with a position angle (P.A.) of $73.5\degr$.

    This letter concentrates on the continuum emission and the H30$\alpha$ RRL at 231.90093~GHz. An analysis of the molecular-line data will be presented in a separate paper. In the SPW containing this RRL, centred on 230.0~GHz, we subtracted the continuum emission with the task \texttt{uvcontsub}. We noticed that adequately flat spectral baselines at the position of  GGD27-MM1 were obtained only when the phase centre of the visibilities was shifted at the position of the source, with the task \texttt{phaseshift}, before continuum subtraction. The final cube of H30$\alpha$ emission was obtained with a spectral resolution of 5 km s$^{-1}$, applying natural weighting of visibilities, and removing all baselines with projected lengths of $<500$~k$\lambda$. In this cube, the synthesised beam has $\theta_s= 60~\mathrm{mas}\times48~\mathrm{mas}$ and P.A. $=80.2\degr$. The continuum and line images were corrected by the response of the primary beam.

%%%%%%%%%%%%%%%%%%%%%%%%%%%%%%%%%%%%%%%%%%%%%%%%%%%%%%%%%%%%%%
\section{Results and discussion}

    \subsection{1.3 mm continuum emission}\label{sect:continuum}
    %%%%%%%%%%%%%%%%%%%%%%%
    %% Continuum results %%
    %%%%%%%%%%%%%%%%%%%%%%%

\begin{figure*}[!htb]
        \centering
        \includegraphics[width=0.75\textwidth]{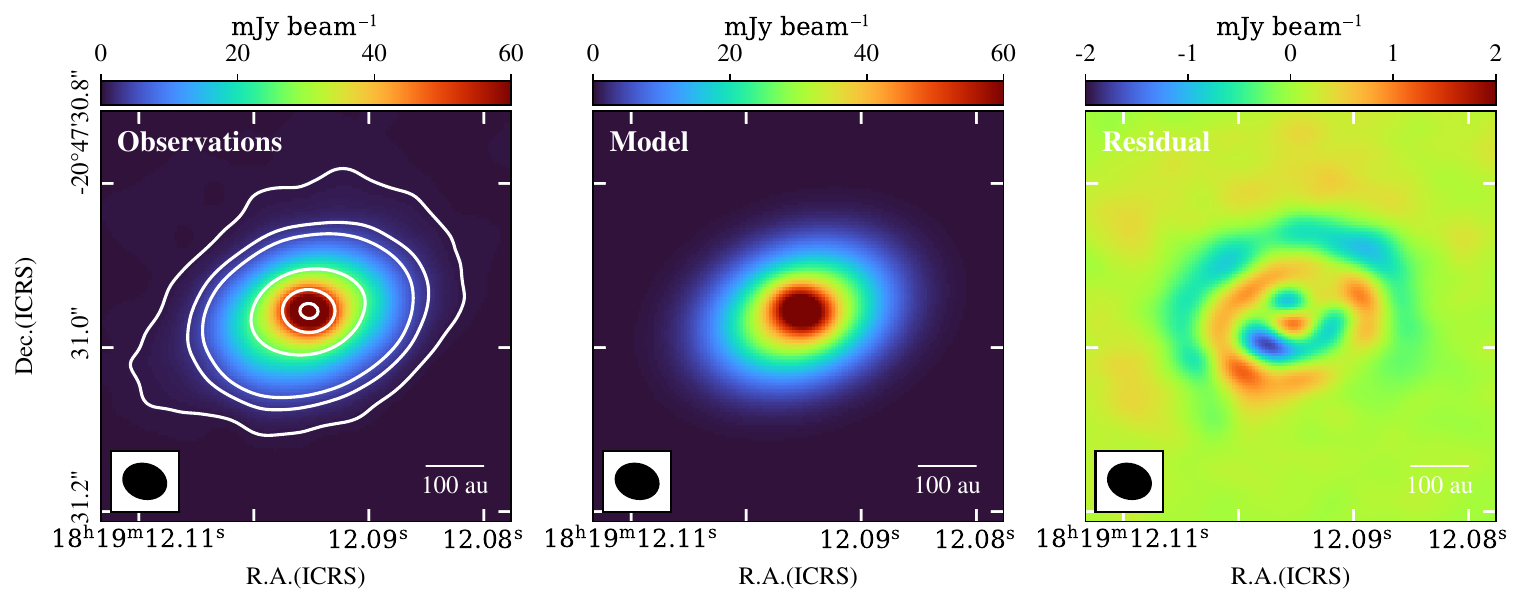}
        
        \caption{\textit{Left:} ALMA band 6 continuum map at 1.3~mm (226.2~GHz). The contour levels are $(10,50,100,500,1000,1400)\times\sigma$, with $\sigma=0.046~\mathrm{mJy/beam}$. \textit{Middle:} Two-component 2D Gaussian best model fit. \textit{Right:} Residual of the two-component 2D Gaussian fit. The beam size ($\theta=53~\mathrm{mas}\times42~\mathrm{mas}$) is represented by the black ellipse at the bottom left of each panel.}
        \label{fig:cont_mod_res}
    \end{figure*}

    The 1.3 mm continuum emission from the high-mass protostar IRAS18162$-$2048 (a.k.a. GGD27-MM1) is shown in the left panel of Fig.~\ref{fig:cont_mod_res}. The source exhibits a smooth elliptical morphology with a prominent central peak that clearly deviates from a single 2D Gaussian intensity distribution. A similar morphology was previously reported by \citet{girart2018} in their 1.14 mm observations. These authors removed the compact central source in the visibility domain and modelled the remaining emission using a double power-law disk, arguing that a single Gaussian profile could not adequately reproduce the observed brightness distribution.
    
    Indeed, fitting the continuum emission with a single 2D Gaussian leaves significant residuals at the centre of the source, most noticeably at the 1000 and 1400$\sigma$ contour levels shown in the left panel of Fig.~\ref{fig:cont_mod_res}. We therefore modelled the continuum emission using two 2D Gaussian components fitted simultaneously using the CASA task \texttt{imfit} without constraining any parameter (middle and right panels of Fig~\ref{fig:cont_mod_res}), to reproduce both the broad and narrow emission. The parameters discussed below correspond to beam-deconvolved values.
    
    The broad Gaussian component successfully reproduces the overall extended morphology of the GGD27-MM1 disk. The best-fit model yields FWHM sizes of $141.8\pm0.7$ mas ($198.5\pm1.0$ au) and $96.0\pm0.5$ mas ($134.4\pm0.7$ au) along the major and minor axes, respectively, with a P.A. of $111.2\pm0.5\degr$. These dimensions imply a disk inclination of $47.3\pm0.9\degr$. The integrated flux density of the broad component is $334.4\pm1.5$ mJy. All these properties are consistent with the disk model derived by \citet{girart2018}, supporting the interpretation that this component traces the dusty circumstellar disk surrounding GGD27-MM1.
    
    The narrow Gaussian component reproduces the bright compact emission at the centre of the disk. The fit yields deconvolved FWHM sizes of $13.8\pm2.4$ mas ($19.3\pm3.4$ au) and $9\pm4$ mas ($12.6\pm5.6$ au) along the major and minor axes, respectively, with a P.A. of $39\pm24\degr$. The integrated flux density of this component is $31.0\pm0.4$ mJy. The complete set of fitted parameters is listed in Table~\ref{tab:continuum_fit}. Interestingly, the P.A. of the narrow component differs significantly from that of the broad disk component. Despite the relatively large uncertainty, the best-fit P.A. of the compact emission is consistent with the direction of the radio jet \citep[P.A.$_{\rm jet}\sim20$--$30\degr$;][]{marti1995,marti1998} rather than with the major axis of the disk \citep[P.A.$_{\rm disk}\sim110$--$114\degr$;][this work]{girart2018}. This suggests that the compact continuum component may be tracing a distinct physical structure, potentially associated with the innermost ionised jet.
    
    Finally, we used the in-band continuum measurements to estimate the spectral index across the 1.3 mm band (see Appendix~\ref{sect:inband_alpha} and Table~\ref{tab:spectral_index}). A linear fit in logarithmic space yields a spectral index of $\alpha = 2.22\pm0.15$ (Fig.~\ref{fig:spectral_index}), indicative of nearly optically thick dust emission. This value is in excellent agreement with that reported by \citet{girart2018}, who derived $\alpha \simeq 2.2$. A detailed physical model of the dusty disk can be found in \citet{anez2020}.
    
    %%%%%%%%%%%%%%%%%%%%%%%%%%%%%%%%%
    %% H30a results - line profile %%
    %%%%%%%%%%%%%%%%%%%%%%%%%%%%%%%%%

    \subsection{The H30$\alpha$ detection}\label{sect:h30a}

    Our data cubes show emission around the frequency of the H30$\alpha$ line (rest frequency $=231.90093$~GHz). However, it is important to consider that some molecular transitions could contaminate the spectrum, hampering the identification of this RRL. For instance, \citet{sanchez-monge2025} noted that the main potential contaminant in the vicinity of H30$\alpha$ is the group of $^{33}$SO$_2$ ($3_\mathrm{9,K_c}-2_\mathrm{10,K_c}$) transitions between 231.89 and 231.90~GHz, although they estimated that their relative contribution in the protostar W75N(B)-VLA3 was only $\sim15\%$ with respect to H30$\alpha$.

    To evaluate this possible contamination in our data, we obtained two data cubes: one with all visibilities, and another one in which we discarded baselines shorter than $500$~k$\lambda$ (thereby retaining only the most compact emission). Figure~\ref{fig:line_profile_33SO2} compares the spectra from these two cubes, shown in red and black, respectively. On scales of $0.5\arcsec$ (top panel), the spectrum clearly shows two bright lines in the unfiltered cube. The line at higher frequency is the SO$_2$ ($14_{3,11}-14_{2,12};\, \nu_2=1$) transition at $\sim231.9805$~GHz, while the one close to the H30$\alpha$ frequency has a similar profile and is associated with $^{33}$SO$_2$. However, these molecular features disappear after filtering out the short baselines, indicating that they arise from spatially extended emission. Moreover, both lines show a similar spatio-kinematical pattern in the unfiltered cube when assuming the rest frequencies of the SO$_2$ and $^{33}$SO$_2$ transitions mentioned above, with a trend compatible with a rotating structure, as seen in \cite{anez2020} also using SO$_2$ transitions at lower angular resolution. Thus we conclude that most of the emission of the second line actually correspond to extended $^{33}$SO$_2$ emission.

    In contrast, when the spectrum is extracted at the continuum peak pixel position, two weak wide spectral features remain clearly visible in both the original and filtered cubes, while the narrow molecular lines are absent (bottom panel of Fig.~\ref{fig:line_profile_33SO2}). We propose that these two wide spectral components correspond to RRL emission. In the following, the spectral line data attributed to H30$\alpha$ refer to the cube after filtering our visibilities $<500$~k$\lambda$.
     
\begin{figure}[!htb]
        \centering
        \includegraphics[width=0.50\textwidth]{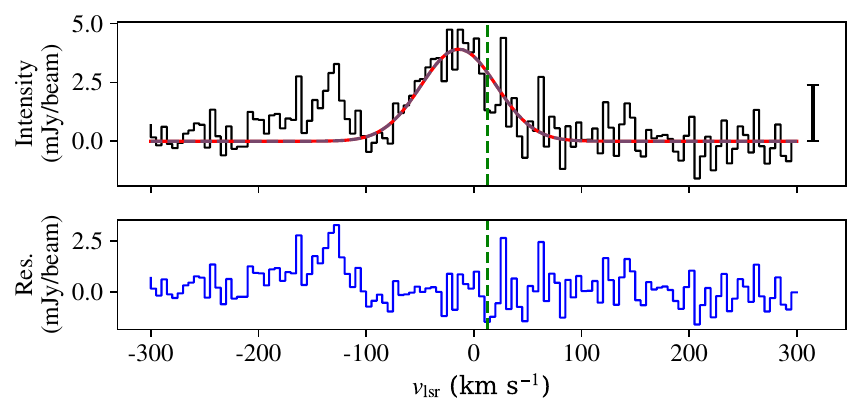}
        
        \caption{\textit{Top:} Spectrum extracted from the continuum peak pixel from the narrow component (see Table~\ref{tab:continuum_fit}) in black and best fit to the main feature in red. \textit{Bottom:} Residual of the Gaussian profile fit. The velocity scale assumes the rest frequency of H30$\alpha$. The dashed green line indicates the systemic velocity of the source ($v_\mathrm{lsr}=+12.5~\mathrm{km~s^{-1}}$). The vertical bar to the right represents the $3\sigma$ confidence level, where $\sigma=0.8$~mJy/beam.}
        \label{fig:line_profile}
    \end{figure}

    We fitted the spectrum shown in Fig.~\ref{fig:line_profile} with a Gaussian profile. The resulting parameters of this fit are a FWHM of $82\pm8~\mathrm{km~s^{-1}}$, a peak intensity of $3.9\pm0.3~\mathrm{mJy~beam^{-1}}$, and a radial velocity with respect to the local standard of rest (lsr) of $v_\mathrm{lsr}=-14\pm3~\mathrm{km~s^{-1}}$, significantly offset from the systemic velocity of the source ($+12.5~\mathrm{km~s^{-1}}$; \citealt{anez2020}). There is a secondary component peaking at a radial velocity of $v_\mathrm{lsr}\sim-127~\mathrm{km~s^{-1}}$. This feature is weaker but clearly detected, with a peak intensity approximately 50\% that of the main feature. The velocity separation between the two peaks is $113\pm6~\mathrm{km~s^{-1}}$. At first glance, this velocity offset may be compatible with the expected separation between H$\,n\alpha$ and He$\,n\alpha$ recombination lines ($\sim-122~\mathrm{km~s^{-1}}$; \citealt{estalella1999}). However, He$\,n\alpha$ lines are typically only $\sim10\%$ as bright as their hydrogen counterparts \citep{zhang2019G45}, substantially lower than the observed intensity ratio. We therefore favour an interpretation in which both features correspond to distinct velocity components of the H30$\alpha$ line \citep[e.g.][]{zhang2019G45}. Under this assumption, both components are blueshifted with respect to the systemic velocity. Nonetheless, due to the weak nature of this feature, we focus on the main emission line as H30$\alpha$ candidate. See Appendix~\ref{sect:second_peak} for more on this second peak.

    To assess whether the H30$\alpha$ emission is consistent with a thermal origin, we estimated the ratio between the RRL and free-free continuum flux densities ($S_L/S_C$). Equation~5 of \citet{sanchez-monge2025} (see also Eq.~23 of \citealt{anglada2018}) provides the expected $S_L/S_C$ ratio for a biconical thermal jet, assuming an ionised helium-to-hydrogen abundance ratio of 0.1. Since the 1.3~mm continuum in our images is dominated by dust emission, we estimated the free-free contribution by extrapolating from the 1.3 cm flux density reported by \citet{carrasco-gonzalez2012}, $1.7\pm0.2$~mJy, assuming a free-free spectral index of $\alpha=+0.6$. This yields an expected free-free continuum flux density of $6.8\pm0.8$ mJy at $\sim1.3$~mm. For $T_e=10^4$~K and $\Delta v = 82\pm8~\mathrm{km\,s^{-1}}$, the predicted thermal H30$\alpha$ flux density is $4.7\pm0.7$ mJy. This value is in excellent agreement with the observed peak intensity, supporting a thermal origin for the detected H30$\alpha$ emission.

\begin{figure}[!htb]
        \centering
        \includegraphics[width=0.32\textwidth]{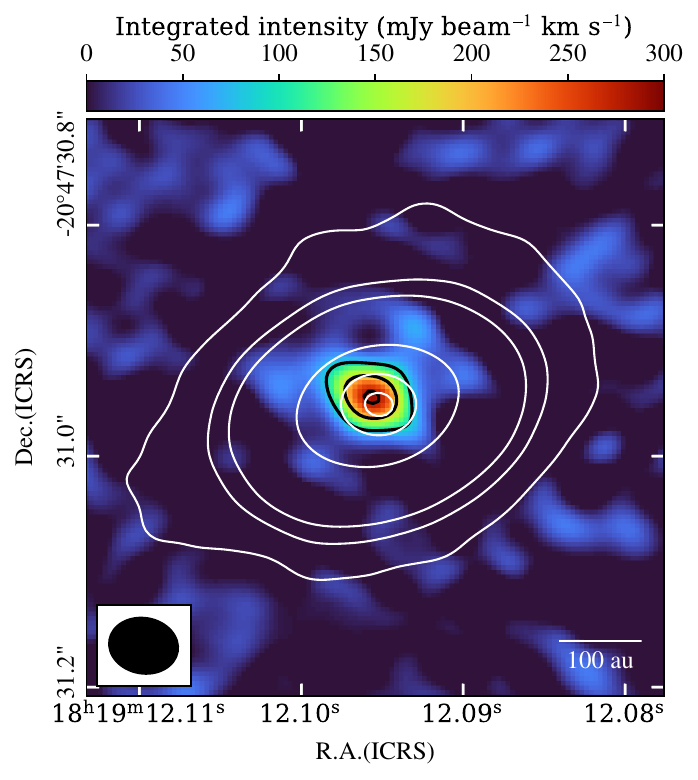}
        
        \caption{
        Moment 0 map for the H30$\alpha$ line integrated in the velocity range $-75\leq v_\mathrm{lsr}~(\mathrm{km~s^{-1}})\leq+20$. The thick black contours are $(5,10,15)\times\sigma_\mathrm{mom0}$, where $\sigma_\mathrm{mom0}=18.4~\mathrm{mJy~beam^{-1}~km~s^{-1}}$. The white contours are the same as in Fig.~\ref{fig:cont_mod_res}.}
        \label{fig:mom0}
    \end{figure}

\begin{figure}[!htb]
        \centering
        \includegraphics[width=0.40
        \textwidth]{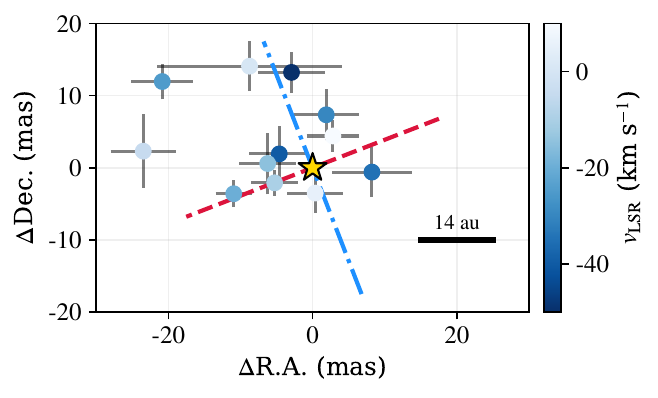}
        
        \caption{Centroid map for the H30$\alpha$ line in the velocity range $-75\leq v_\mathrm{lsr}~(\mathrm{km~s^{-1}})\leq+20$. The dashed red line indicates the disk P.A. from the broad component of the continuum fit ($111.2\degr$; see Table~\ref{tab:continuum_fit}), whereas the dash-dotted  blue line is perpendicular to the red line and represents the jet P.A. ($21.2\degr$).}
        \label{fig:centroids}
    \end{figure}

    To further verify that the H30$\alpha$ emission is associated with the compact narrow continuum component, we constructed a moment 0 map of the filtered cube, integrating in the range $-75 \leq v_\mathrm{lsr}~(\mathrm{km~s^{-1}}) \leq +20$, to cover the main spectral component exclusively. Figure~\ref{fig:mom0} presents the resulting H30$\alpha$ integrated intensity map. The emission is unresolved and confined to a region comparable in size to the synthesised beam, indicating that the ionised gas originates from a region smaller than $\sim$75~au. Nevertheless, the emission exhibits a possible marginal elongation towards the blueshifted side of the jet axis (i.e. northeast), and the peak of the moment 0 map appears slightly displaced to the northeast with respect to the continuum peak, in the direction of the blueshifted lobe of the CO outflow. This can also be seen in the map of emission centroids in each velocity channel (Fig. \ref{fig:centroids}), where there is a clear bias towards the northeast.
    Although these features remain below the angular resolution required for a robust morphological analysis, they may hint at a contribution from ionised gas associated with the jet.

    The centroid distribution shown in Fig.~\ref{fig:centroids} further constrains the size of the emitting region to $\leq 50$~au, consistent with the deconvolved size of the continuum narrow component. We built this map by fitting elliptical Gaussians with the CASA task \texttt{imfit} to the emission in individual channels over the velocity range $-75\leq v_\mathrm{lsr}~(\mathrm{km~s^{-1}})\leq+20$, considering only channels whose fitted peak intensities exceeded five times their corresponding fitting uncertainties. While the compactness of the emission and its broad linewidth are roughly consistent with those of hypercompact or very young \ion{H}{ii} regions \citep{kim2018}, several observational properties argue against this interpretation. In particular, the measured linewidth of $\sim82~\mathrm{km~s^{-1}}$ is somewhat larger than typically expected for a compact \ion{H}{ii} region, and the line centroid is significantly blueshifted from the systemic velocity of the cloud. If the recombination line were primarily tracing an embedded \ion{H}{ii} region, one would instead expect the emission to be centred close to the systemic velocity and show both redshifted and blueshifted emission. The combination of the large velocity offset, the predominance of blueshifted emission, and the apparent alignment with the jet axis strongly suggests that the H30$\alpha$ emission traces ionised gas associated with the approaching side of the radio jet. In this scenario, the redshifted counterpart may be obscured by the optically thick dusty disk discussed in Sect.~\ref{sect:continuum}.

    In summary, we report the first detection of thermal emission of the H30$\alpha$ RRL towards GGD27-MM1. The spectral properties, compact morphology, and centroid distribution collectively indicate that the ionised gas is associated with the jet powering the parsec-scale HH80-81 outflow. Higher-sensitivity observations at an improved angular resolution, together with observations of multiple RRL transitions, will be essential to fully characterise the origin, kinematics, and physical conditions of the ionised gas in GGD27-MM1. This result, together with the recent detection of thermal RRL in W75N(B) \citep{sanchez-monge2025}, opens a new window for future detailed studies of the 3D kinematics (radial velocities and proper motions) and physical properties of protostellar ionised jets.

%%%%%%%%%%%%%%%%%%%%%%%%%%%%%%%%%%%%%%%%%%%%%%%%%%%%%%%%%%%%%%

\begin{acknowledgements}
    We thank the referee for their constructive comments. R.F., J.F.G., M.O., G.A., A.F.P.M., and G.B.C. acknowledge support from the grants PID2023-146295NB-I00, and from the Severo Ochoa grant CEX2021-001131-S funded by MCIN/AEI/ 10.13039/501100011033 and ERDF/EU. C.Y.L. acknowledges financial support through the INAF Large Grant The role of MAGnetic fields in MAssive star formation (MAGMA). This paper makes use of the following ALMA data: ADS/JAO.ALMA\#2016.1.01036.S ALMA is a partnership of ESO (representing its member states), NSF (USA) and NINS (Japan), together with NRC (Canada), NSTC and ASIAA (Taiwan), and KASI (Republic of Korea), in cooperation with the Republic of Chile. The Joint ALMA Observatory is operated by ESO, AUI/NRAO and NAOJ.
\end{acknowledgements}

%%%%%%%%%%%%%%%%%%%%%%%%%%%%%%%%%%%%%%%%%%%%%%%%%%%%%%%%%%%%%%

\bibliographystyle{aa}
\bibliography{phd_bibliography.bib}

@ARTICLE{anglada2018,
	author = {{Anglada}, G. and {Rodr{\'{\i}}guez}, L.~F. and {Carrasco-Gonz{\'a}lez}, C.
	},
	title = "{Radio jets from young stellar objects}",
	journal = {\aapr},
	archivePrefix = "arXiv",
	eprint = {1806.06444},
	primaryClass = "astro-ph.SR",
	year = 2018,
	month = jun,
	volume = 26,
	eid = {3},
	pages = {3},
	doi = {10.1007/s00159-018-0107-z},
	adsurl = {http://adsabs.harvard.edu/abs/2018A%26ARv..26....3A}
}

@ARTICLE{anez2020,
       author = {{A{\~n}ez-L{\'o}pez}, N. and {Osorio}, M. and {Busquet}, G. and {Girart}, J.~M. and {Mac{\'\i}as}, E. and {Carrasco-Gonz{\'a}lez}, C. and {Curiel}, S. and {Estalella}, R. and {Fern{\'a}ndez-L{\'o}pez}, M. and {Galv{\'a}n-Madrid}, R. and {Kwon}, J. and {Torrelles}, J.~M.},
        title = "{Modeling the Accretion Disk around the High-mass Protostar GGD 27-MM1}",
      journal = {\apj},
         year = 2020,
        month = jan,
       volume = {888},
       number = {1},
          eid = {41},
        pages = {41},
          doi = {10.3847/1538-4357/ab5dbc},
archivePrefix = {arXiv},
       eprint = {1911.12398},
 primaryClass = {astro-ph.SR},
       adsurl = {https://ui.adsabs.harvard.edu/abs/2020ApJ...888...41A}
}

@ARTICLE{bally2016,
	author = {{Bally}, J.},
	title = "{Protostellar Outflows}",
	journal = {\araa},
	year = 2016,
	month = sep,
	volume = 54,
	pages = {491-528},
	doi = {10.1146/annurev-astro-081915-023341},
	adsurl = {http://adsabs.harvard.edu/abs/2016ARA%26A..54..491B}
}

@ARTICLE{bally2023,
       author = {{Bally}, John and {Reipurth}, Bo},
        title = "{HH 80/81: Structure and Kinematics of the Fastest Protostellar Outflow}",
      journal = {\apj},
         year = 2023,
        month = nov,
       volume = {958},
       number = {1},
          eid = {99},
        pages = {99},
          doi = {10.3847/1538-4357/acf028},
archivePrefix = {arXiv},
       eprint = {2308.13638},
 primaryClass = {astro-ph.GA},
       adsurl = {https://ui.adsabs.harvard.edu/abs/2023ApJ...958...99B}
}

@ARTICLE{beuther2025,
       author = {{Beuther}, H. and {Kuiper}, R. and {Tafalla}, M.},
        title = "{Star Formation from Low to High Mass: A Comparative View}",
      journal = {\araa},
         year = 2025,
        month = aug,
       volume = {63},
       number = {1},
        pages = {1-44},
          doi = {10.1146/annurev-astro-013125-122023},
archivePrefix = {arXiv},
       eprint = {2501.16866},
 primaryClass = {astro-ph.GA},
       adsurl = {https://ui.adsabs.harvard.edu/abs/2025ARA&A..63....1B}
}

@PHDTHESIS{briggs1995,
       author = {{Briggs}, Daniel Shenon},
        title = "{High fidelity deconvolution of moderately resolved sources}",
       school = {New Mexico Institute of Mining and Technology},
         year = 1995,
        month = jan,
       adsurl = {https://ui.adsabs.harvard.edu/abs/1995PhDT.......238B}
}

@ARTICLE{brown1978,
       author = {{Brown}, R.~L. and {Lockman}, F.~J. and {Knapp}, G.~R.},
        title = "{Radio recombination lines.}",
      journal = {\araa},
         year = 1978,
        month = jan,
       volume = {16},
        pages = {445-485},
          doi = {10.1146/annurev.aa.16.090178.002305},
       adsurl = {https://ui.adsabs.harvard.edu/abs/1978ARA&A..16..445B}
}

@ARTICLE{carrasco-gonzalez2012,
       author = {{Carrasco-Gonz{\'a}lez}, Carlos and {Galv{\'a}n-Madrid}, Roberto and {Anglada}, Guillem and {Osorio}, Mayra and {D'Alessio}, Paola and {Hofner}, Peter and {Rodr{\'\i}guez}, Luis F. and {Linz}, Hendrik and {Araya}, Esteban D.},
        title = "{Resolving the Circumstellar Disk around the Massive Protostar Driving the HH 80-81 Jet}",
      journal = {\apjl},
         year = 2012,
        month = jun,
       volume = {752},
       number = {2},
          eid = {L29},
        pages = {L29},
          doi = {10.1088/2041-8205/752/2/L29},
archivePrefix = {arXiv},
       eprint = {1205.3302},
 primaryClass = {astro-ph.GA},
       adsurl = {https://ui.adsabs.harvard.edu/abs/2012ApJ...752L..29C}
}

@ARTICLE{dupree1970,
       author = {{Dupree}, A.~K. and {Goldberg}, Leo},
        title = "{Radiofrequency Recombination Lines}",
      journal = {\araa},
         year = 1970,
        month = jan,
       volume = {8},
        pages = {231},
          doi = {10.1146/annurev.aa.08.090170.001311},
       adsurl = {https://ui.adsabs.harvard.edu/abs/1970ARA&A...8..231D}
}

@book{estalella1999,
  author       = {Estalella, Robert and Anglada, Guillem},
  title        = {Introducci{\'o}n a la F{\'i}sica del Medio Interestelar},
  year         = {1999},
  publisher    = {Edicions UB},
  address      = {Barcelona, España},
  series       = {Textos Docents},
  number       = {50},
  pages        = {140},
  isbn         = {84-8338-098-6}
}

@ARTICLE{fedriani2026,
       author = {{Fedriani}, R. and {Anglada}, G. and {o Garatti}, A. Caratti and {G{\'o}mez}, J.~F. and {Masqu{\'e}}, J. and {Osorio}, M. and {Stecklum}, B. and {Rodr{\'\i}guez-Kamenetzky}, A.~R. and {Galv{\'a}n-Madrid}, R. and {Carrasco-Gonz{\'a}lez}, C. and {Bl{\'a}zquez-Calero}, G. and {Placinta-Mitrea}, A.~F. and {Sanna}, A. and {Cesaroni}, R. and {Moscadelli}, L. and {Ray}, T.~P. and {Coffey}, D. and {Fuller}, G.~A.},
        title = "{Diverse stages of star formation in the IRAS 18162-2048 region: Emergence of UV feedback}",
      journal = {\aap},
         year = 2026,
        month = mar,
       volume = {708},
          eid = {A11},
        pages = {A11},
          doi = {10.1051/0004-6361/202558460},
archivePrefix = {arXiv},
       eprint = {2512.07604},
 primaryClass = {astro-ph.GA},
       adsurl = {https://ui.adsabs.harvard.edu/abs/2026A&A...708A..11F}
}

@ARTICLE{fernandez-lopez2011dust,
       author = {{Fern{\'a}ndez-L{\'o}pez}, M. and {Curiel}, S. and {Girart}, J.~M. and {Ho}, P.~T.~P. and {Patel}, N. and {G{\'o}mez}, Y.},
        title = "{Millimeter and Submillimeter High Angular Resolution Interferometric Observations: Dust in the Heart of IRAS 18162-2048}",
      journal = {\aj},
         year = 2011,
        month = mar,
       volume = {141},
       number = {3},
          eid = {72},
        pages = {72},
          doi = {10.1088/0004-6256/141/3/72},
archivePrefix = {arXiv},
       eprint = {1011.2237},
 primaryClass = {astro-ph.SR},
       adsurl = {https://ui.adsabs.harvard.edu/abs/2011AJ....141...72F}
}

@ARTICLE{fernandez-lopez2011molecule,
       author = {{Fern{\'a}ndez-L{\'o}pez}, M. and {Girart}, J.~M. and {Curiel}, S. and {G{\'o}mez}, Y. and {Ho}, P.~T.~P. and {Patel}, N.},
        title = "{A Rotating Molecular Disk Toward IRAS 18162-2048, the Exciting Source of HH 80-81}",
      journal = {\aj},
         year = 2011,
        month = oct,
       volume = {142},
       number = {4},
          eid = {97},
        pages = {97},
          doi = {10.1088/0004-6256/142/4/97},
archivePrefix = {arXiv},
       eprint = {1107.3176},
 primaryClass = {astro-ph.GA},
       adsurl = {https://ui.adsabs.harvard.edu/abs/2011AJ....142...97F}
}

@ARTICLE{francis2020,
       author = {{Francis}, Logan and {Johnstone}, Doug and {Herczeg}, Gregory and {Hunter}, Todd R. and {Harsono}, Daniel},
        title = "{On the Accuracy of the ALMA Flux Calibration in the Time Domain and across Spectral Windows}",
      journal = {\aj},
         year = 2020,
        month = dec,
       volume = {160},
       number = {6},
          eid = {270},
        pages = {270},
          doi = {10.3847/1538-3881/abbe1a},
archivePrefix = {arXiv},
       eprint = {2010.02186},
 primaryClass = {astro-ph.IM},
       adsurl = {https://ui.adsabs.harvard.edu/abs/2020AJ....160..270F}
}

@ARTICLE{girart2017,
       author = {{Girart}, J.~M. and {Estalella}, R. and {Fern{\'a}ndez-L{\'o}pez}, M. and {Curiel}, S. and {Frau}, P. and {Galvan-Madrid}, R. and {Rao}, R. and {Busquet}, G. and {Ju{\'a}rez}, C.},
        title = "{The Circumestellar Disk of the B0 Protostar Powering the HH 80-81 Radio Jet}",
      journal = {\apj},
         year = 2017,
        month = sep,
       volume = {847},
       number = {1},
          eid = {58},
        pages = {58},
          doi = {10.3847/1538-4357/aa81c9},
archivePrefix = {arXiv},
       eprint = {1707.06463},
 primaryClass = {astro-ph.SR},
       adsurl = {https://ui.adsabs.harvard.edu/abs/2017ApJ...847...58G}
}

@ARTICLE{girart2018,
	author = {{Girart}, J.~M. and {Fern{\'a}ndez-L{\'o}pez}, M. and {Li}, Z.-Y. and 
	{Yang}, H. and {Estalella}, R. and {Anglada}, G. and {{\'A}{\~n}ez-L{\'o}pez}, N. and 
	{Busquet}, G. and {Carrasco-Gonz{\'a}lez}, C. and {Curiel}, S. and 
	{Galvan-Madrid}, R. and {G{\'o}mez}, J.~F. and {de Gregorio-Monsalvo}, I. and 
	{Jim{\'e}nez-Serra}, I. and {Krasnopolsky}, R. and {Mart{\'{\i}}}, J. and 
	{Osorio}, M. and {Padovani}, M. and {Rao}, R. and {Rodr{\'{\i}}guez}, L.~F. and 
	{Torrelles}, J.~M.},
	title = "{Resolving the Polarized Dust Emission of the Disk around the Massive Star Powering the HH 80-81 Radio Jet}",
	journal = {\apjl},
	archivePrefix = "arXiv",
	eprint = {1803.06165},
	primaryClass = "astro-ph.SR",
	year = 2018,
	month = apr,
	volume = 856,
	eid = {L27},
	pages = {L27},
	doi = {10.3847/2041-8213/aab76b},
	adsurl = {http://adsabs.harvard.edu/abs/2018ApJ...856L..27G}
}

@ARTICLE{gomez2003,
       author = {{G{\'o}mez}, Y. and {Rodr{\'\i}guez}, L.~F. and {Girart}, J.~M. and {Garay}, G. and {Mart{\'\i}}, J.},
        title = "{VLA and BIMA Observations toward the Exciting Source of the Massive HH 80-81 Outflow}",
      journal = {\apj},
         year = 2003,
        month = nov,
       volume = {597},
       number = {1},
        pages = {414-423},
          doi = {10.1086/378403},
archivePrefix = {arXiv},
       eprint = {astro-ph/0307511},
 primaryClass = {astro-ph},
       adsurl = {https://ui.adsabs.harvard.edu/abs/2003ApJ...597..414G}
}

@ARTICLE{gyulbudaghian1978,
       author = {{Gyulbudaghian}, A.~L. and {Glushkov}, Yu. I. and {Denisyuk}, E.~K.},
        title = "{New Herbig-Haro objects.}",
      journal = {\apjl},
         year = 1978,
        month = sep,
       volume = {224},
        pages = {L137-L138},
          doi = {10.1086/182777},
       adsurl = {https://ui.adsabs.harvard.edu/abs/1978ApJ...224L.137G}
}

@ARTICLE{jimenez-serra2011,
       author = {{Jim{\'e}nez-Serra}, I. and {Mart{\'\i}n-Pintado}, J. and {B{\'a}ez-Rubio}, A. and {Patel}, N. and {Thum}, C.},
        title = "{Extremely Broad Radio Recombination Maser Lines Toward the High-velocity Ionized Jet in Cepheus A HW2}",
      journal = {\apjl},
         year = 2011,
        month = may,
       volume = {732},
       number = {2},
          eid = {L27},
        pages = {L27},
          doi = {10.1088/2041-8205/732/2/L27},
archivePrefix = {arXiv},
       eprint = {1103.4073},
 primaryClass = {astro-ph.SR},
       adsurl = {https://ui.adsabs.harvard.edu/abs/2011ApJ...732L..27J}
}

@ARTICLE{kim2018,
       author = {{Kim}, W.-J. and {Urquhart}, J.~S. and {Wyrowski}, F. and {Menten}, K.~M. and {Csengeri}, T.},
        title = "{New detections of (sub)millimeter hydrogen radio recombination lines towards high-mass star-forming clumps}",
      journal = {\aap},
         year = 2018,
        month = aug,
       volume = {616},
          eid = {A107},
        pages = {A107},
          doi = {10.1051/0004-6361/201732330},
archivePrefix = {arXiv},
       eprint = {1804.06999},
 primaryClass = {astro-ph.GA},
       adsurl = {https://ui.adsabs.harvard.edu/abs/2018A&A...616A.107K}
}

@ARTICLE{law2022,
       author = {{Law}, Chi-Yan and {Tan}, Jonathan C. and {Gorai}, Prasanta and {Zhang}, Yichen and {Fedriani}, Rub{\'e}n and {Tafoya}, Daniel and {Tanaka}, Kei E.~I. and {Cosentino}, Giuliana and {Yang}, Yao-Lun and {Mardones}, Diego and {Beltr{\'a}n}, Maria T. and {Garay}, Guido},
        title = "{Isolated Massive Star Formation in G28.20-0.05}",
      journal = {\apj},
         year = 2022,
        month = nov,
       volume = {939},
       number = {2},
          eid = {120},
        pages = {120},
          doi = {10.3847/1538-4357/ac90c7},
archivePrefix = {arXiv},
       eprint = {2201.01411},
 primaryClass = {astro-ph.GA},
       adsurl = {https://ui.adsabs.harvard.edu/abs/2022ApJ...939..120L}
}

@ARTICLE{marti1993,
       author = {{Marti}, J. and {Rodriguez}, L.~F. and {Reipurth}, B.},
        title = "{HH 80-81: A Highly Collimated Herbig-Haro Complex Powered by a Massive Young Star}",
      journal = {\apj},
         year = 1993,
        month = oct,
       volume = {416},
        pages = {208},
          doi = {10.1086/173227},
       adsurl = {https://ui.adsabs.harvard.edu/abs/1993ApJ...416..208M}
}

@ARTICLE{marti1995,
       author = {{Marti}, J. and {Rodriguez}, L.~F. and {Reipurth}, Bo},
        title = "{Large Proper Motions and Ejection of New Condensations in the HH 80-81 Thermal Radio Jet}",
      journal = {\apj},
         year = "1995",
        month = "Aug",
       volume = {449},
        pages = {184},
          doi = {10.1086/176044},
       adsurl = {https://ui.adsabs.harvard.edu/abs/1995ApJ...449..184M}
}

@ARTICLE{marti1998,
       author = {{Mart{\'\i}}, Josep and {Rodr{\'\i}guez}, Luis F. and {Reipurth}, Bo},
        title = "{Proper Motions of the Inner Condensations in the HH 80-81 Thermal Radio Jet}",
      journal = {\apj},
         year = 1998,
        month = jul,
       volume = {502},
       number = {1},
        pages = {337-341},
          doi = {10.1086/305900},
       adsurl = {https://ui.adsabs.harvard.edu/abs/1998ApJ...502..337M}
}

@ARTICLE{martinez-henares2024,
       author = {{Mart{\'\i}nez-Henares}, Antonio and {Zhang}, Qizhou and {Jim{\'e}nez-Serra}, Izaskun and {Mart{\'\i}n-Pintado}, Jes{\'u}s and {Hu{\'e}lamo}, Nuria and {Prasad}, Sirina and {Moran}, James and {B{\'a}ez-Rubio}, Alejandro},
        title = "{Imaging the Jet of MWC 349A with Resolved Radio Recombination Line Emission from ALMA}",
      journal = {\apjl},
         year = 2024,
        month = aug,
       volume = {971},
       number = {1},
          eid = {L7},
        pages = {L7},
          doi = {10.3847/2041-8213/ad638d},
archivePrefix = {arXiv},
       eprint = {2407.13681},
 primaryClass = {astro-ph.SR},
       adsurl = {https://ui.adsabs.harvard.edu/abs/2024ApJ...971L...7M}
}

@ARTICLE{masque2012,
       author = {{Masqu{\'e}}, Josep M. and {Girart}, Josep M. and {Estalella}, Robert and {Rodr{\'\i}guez}, Luis F. and {Beltr{\'a}n}, Maria T.},
        title = "{Centimeter Continuum Observations of the Northern Head of the HH 80/81/80N Jet: Revising the Actual Dimensions of a Parsec-scale Jet}",
      journal = {\apjl},
         year = 2012,
        month = oct,
       volume = {758},
       number = {1},
          eid = {L10},
        pages = {L10},
          doi = {10.1088/2041-8205/758/1/L10},
archivePrefix = {arXiv},
       eprint = {1209.1254},
 primaryClass = {astro-ph.SR},
       adsurl = {https://ui.adsabs.harvard.edu/abs/2012ApJ...758L..10M}
}

@INPROCEEDINGS{mcmullin2007,
       author = {{McMullin}, J.~P. and {Waters}, B. and {Schiebel}, D. and {Young}, W. and {Golap}, K.},
        title = "{CASA Architecture and Applications}",
    booktitle = {Astronomical Data Analysis Software and Systems XVI},
         year = 2007,
       editor = {{Shaw}, R.~A. and {Hill}, F. and {Bell}, D.~J.},
       series = {Astronomical Society of the Pacific Conference Series},
       volume = {376},
        month = oct,
        pages = {127},
       adsurl = {https://ui.adsabs.harvard.edu/abs/2007ASPC..376..127M}
}

@ARTICLE{mohan2023obs,
       author = {{Mohan}, Sreelekshmi and {Vig}, S. and {Varricatt}, W.~P. and {Tej}, A.},
        title = "{Imaging of HH80-81 Jet in the Near-infrared Shock Tracers H$_{2}$ and [Fe II]}",
      journal = {\apj},
         year = 2023,
        month = jan,
       volume = {942},
       number = {2},
          eid = {76},
        pages = {76},
          doi = {10.3847/1538-4357/aca413},
archivePrefix = {arXiv},
       eprint = {2211.08012},
 primaryClass = {astro-ph.SR},
       adsurl = {https://ui.adsabs.harvard.edu/abs/2023ApJ...942...76M}
}

@ARTICLE{qiu2009,
       author = {{Qiu}, Keping and {Zhang}, Qizhou},
        title = "{Discovery of Extremely High Velocity ``Molecular Bullets'' in the HH 80-81 High-Mass Star-Forming Region}",
      journal = {\apjl},
         year = 2009,
        month = sep,
       volume = {702},
       number = {1},
        pages = {L66-L71},
          doi = {10.1088/0004-637X/702/1/L66},
archivePrefix = {arXiv},
       eprint = {0907.5040},
 primaryClass = {astro-ph.GA},
       adsurl = {https://ui.adsabs.harvard.edu/abs/2009ApJ...702L..66Q}
}

@ARTICLE{reifenstein1970,
       author = {{Reifenstein}, E.~C. and {Wilson}, T.~L. and {Burke}, B.~F. and {Mezger}, P.~G. and {Altenhoff}, W.~J.},
        title = "{A Survey of H1O9{\ensuremath{\alpha}} Recombination Line Emission in Galactic HII Regions of the Northern Sky}",
      journal = {\aap},
         year = 1970,
        month = mar,
       volume = {4},
        pages = {357-377},
       adsurl = {https://ui.adsabs.harvard.edu/abs/1970A&A.....4..357R}
}

@ARTICLE{reipurth1988,
       author = {{Reipurth}, Bo and {Graham}, J.~A.},
        title = "{New Herbig-Haro objects in star-forming regions..}",
      journal = {\aap},
         year = "1988",
        month = "Aug",
       volume = {202},
        pages = {219-239},
       adsurl = {https://ui.adsabs.harvard.edu/abs/1988A&A...202..219R}
}

@ARTICLE{rodriguez1980,
       author = {{Rodriguez}, L.~F. and {Moran}, J.~M. and {Ho}, P.~T.~P. and {Gottlieb}, E.~W.},
        title = "{Radio observations of water vapor, hydroxyl, silicon monoxide, ammonia, carbon monoxide, and compact H II regions in the vicinities of suspected Herbig-Haro objects.}",
      journal = {\apj},
         year = 1980,
        month = feb,
       volume = {235},
        pages = {845-865},
          doi = {10.1086/157687},
       adsurl = {https://ui.adsabs.harvard.edu/abs/1980ApJ...235..845R}
}

@ARTICLE{rodriguez1989,
       author = {{Rodr{\'\i}guez}, L.~F. and {Reipurth}, B.},
        title = "{Detection of radio continuum emission from the Herbig-Haro objects 80 and 81 and their suspected energy source.}",
      journal = {\rmxaa},
         year = "1989",
        month = "May",
       volume = {17},
        pages = {59-63},
       adsurl = {https://ui.adsabs.harvard.edu/abs/1989RMxAA..17...59R}
}

@ARTICLE{sanchez-monge2025,
       author = {{S{\'a}nchez-Monge}, {\'A}. and {G{\'o}mez}, J.~F. and {Torrelles}, J.~M. and {Curiel}, S. and {Girart}, J.~M. and {Surcis}, G. and {Carrasco-Gonz{\'a}lez}, C. and {Anglada}, G. and {Fuller}, G.~A. and {Goddi}, C. and {Vlemmings}, W.~H.~T. and {Rodr{\'\i}guez-Kamenetzky}, A.~R. and {van Langevelde}, H.~J. and {Kim}, J.-S. and {Kim}, S.-W. and {Cant{\'o}}, J.},
        title = "{A compact Keplerian-like disc and H30 {\ensuremath{\alpha}} emission towards the radio jet in the massive protostar W75N(B)-VLA3}",
      journal = {\mnras},
         year = 2025,
        month = oct,
       volume = {543},
       number = {1},
        pages = {662-690},
          doi = {10.1093/mnras/staf1451},
archivePrefix = {arXiv},
       eprint = {2508.21747},
 primaryClass = {astro-ph.GA},
       adsurl = {https://ui.adsabs.harvard.edu/abs/2025MNRAS.543..662S}
}

@ARTICLE{wilson1970,
       author = {{Wilson}, T.~L. and {Mezger}, P.~G. and {Gardner}, F.~F. and {Milne}, D.~K.},
        title = "{A Survey of H 109 {\ensuremath{\alpha}} Recombination Line Emission in Galactic H II Regions of the Southern Sky}",
      journal = {\aap},
         year = 1970,
        month = jul,
       volume = {6},
        pages = {364-384},
       adsurl = {https://ui.adsabs.harvard.edu/abs/1970A&A.....6..364W}
}

@ARTICLE{zhang2019nature,
       author = {{Zhang}, Yichen and {Tan}, Jonathan C. and {Tanaka}, Kei E.~I. and
         {De Buizer}, James M. and {Liu}, Mengyao and {Beltr{\'a}n}, Maria T. and
         {Kratter}, Kaitlin and {Mardones}, Diego and {Garay}, Guido},
        title = "{Dynamics of a massive binary at birth}",
      journal = {Nature Astronomy},
         year = "2019",
        month = "Mar",
       volume = {3},
        pages = {517-523},
          doi = {10.1038/s41550-019-0718-y},
archivePrefix = {arXiv},
       eprint = {1903.07532},
 primaryClass = {astro-ph.SR},
       adsurl = {https://ui.adsabs.harvard.edu/abs/2019NatAs...3..517Z}
}

@ARTICLE{zhang2019G45,
       author = {{Zhang}, Yichen and {Tanaka}, Kei E.~I. and {Rosero}, Viviana and
         {Tan}, Jonathan C. and {Marvil}, Joshua and {Cheng}, Yu and
         {Liu}, Mengyao and {Beltr{\'a}n}, Maria T. and {Garay}, Guido},
        title = "{Discovery of a Photoionized Bipolar Outflow toward the Massive Protostar G45.47+0.05}",
      journal = {\apjl},
         year = "2019",
        month = "Nov",
       volume = {886},
       number = {1},
          eid = {L4},
        pages = {L4},
          doi = {10.3847/2041-8213/ab5309},
archivePrefix = {arXiv},
       eprint = {1910.04271},
 primaryClass = {astro-ph.SR},
       adsurl = {https://ui.adsabs.harvard.edu/abs/2019ApJ...886L...4Z}
}

@ARTICLE{zhang2022,
       author = {{Zhang}, Yichen and {Tanaka}, Kei E.~I. and {Tan}, Jonathan C. and {Yang}, Yao-Lun and {Greco}, Eva and {Beltran}, Maria T. and {Sakai}, Nami and {De Buizer}, James M. and {Rosero}, Viviana and {Fedriani}, Rub{\'e}n and {Garay}, Guido},
        title = "{Massive Protostars in a Protocluster - A Multi-scale ALMA View of G35.20-0.74N.}",
      journal = {\apj},
         year = 2022,
        month = sep,
       volume = {936},
       number = {1},
          eid = {68},
        pages = {68},
          doi = {10.3847/1538-4357/ac847f},
archivePrefix = {arXiv},
       eprint = {2207.11320},
 primaryClass = {astro-ph.GA},
       adsurl = {https://ui.adsabs.harvard.edu/abs/2022ApJ...936...68Z}
}

%%%%%%%%%%%%%%%%%%%%%%%%%%%%%%%%%%%%%%%%%%%%%%%%%%%%%%%%%%%%%%%
% Appendices must be placed after   \end{thebibliography}
% They will be placed automatically on a new page.
%%%%%%%%%%%%%%%%%%%%%%%%%%%%%%%%%%%%%%%%%%%%%%%%%%%%%%%%%%%%%%%

\begin{appendix}
\nolinenumbers
\section{Two-component 2D Gaussian fit results}

    Table~\ref{tab:continuum_fit} summarises the results of the two-component 2D Gaussian fit. The quoted uncertainties in the coordinates are the formal errors in the fit, and should be interpreted as relative errors within the image. The absolute astrometric precision of the continuum image is $\simeq 5$ mas.
    
    \begin{table*}[!htb]
    \caption{Two-component 2D Gaussian fit results to the continuum image of GGD27-MM1.  }
    \label{tab:continuum_fit}      
    \centering          
    \begin{tabular}{l c c c c c c c c}    
    \hline\hline       
    \noalign{\smallskip}
    Comp. & R.A. (ICRS) & Dec. (ICRS) & R.A. error & Dec. error & FWHM$_\mathrm{maj}$ & FWHM$_\mathrm{min}$ & P.A. & $S_\mathrm{1.3mm}$ \\ 
    &(hh:mm:ssssss) &  (dd:mm:sssss) & (mas) & (mas) & (mas) & (mas) & ($\degr$) & (mJy)\\
    \noalign{\smallskip}
    \hline              
    \noalign{\smallskip}
    Broad & 18:19:12.095312& -20:47:30.95837 & 0.25 & 0.18 & $141.8\pm0.7$ & $96.0 \pm0.5$ & $111.2\pm0.5$ & $334.4\pm1.5$\\
    Narrow & 18:19:12.095236 & -20:47:30.95339 & 0.18 & 0.13 & $13.8\pm2.4$ & $9\pm4$ & $39\pm24$ & $31.0\pm0.4$ \\
    \noalign{\smallskip}
    \hline                  
    \end{tabular}
    \tablefoot{The FWHM and P.A. are deconvolved from the beam.}
    \end{table*}

\section{In-band spectral index}\label{sect:inband_alpha}

    We generated four independent continuum images, one for each SPW present in the observation set-up. We measured the flux density over a box of $0.5\arcsec\times0.5\arcsec$ and centred at the continuum peak of the narrow component (see Sect.~\ref{sect:continuum}) for each continuum image. Table~\ref{tab:spectral_index} summarises the frequency of each SPW together with the measured flux density. The uncertainties were estimated from the root mean square (RMS) noise measured in a nearby emission-free region and scaled by the square root of the ratio between the box area and the synthesised beam area.
    
    We derive an in-band spectral index of $2.22\pm0.15$ (Fig.~\ref{fig:spectral_index}). The error is dominated by the uncertainty in the flux density calibration among ALMA SPWs within the same band, which has been estimated to be $\sigma_F=0.8\%$ \citep{francis2020}. Thus, the spectral index uncertainty is $\sigma_\alpha=\sqrt{2}\sigma_F/\ln(\nu_2/\nu_1)$, where $\nu_1$ and $\nu_2$ are the minimum and maximum frequency used in the fit. The spectral index we obtained is consistent with nearly optically thick dust emission and agrees with the spectral index reported by \citet{girart2018}.

    \begin{table}[!htb]
    \caption{Flux density measurements in the four SPW continuum images for GGD27-MM1.}
    \label{tab:spectral_index}      
    \centering          
    \begin{tabular}{l c c}    
    \hline\hline       
    \noalign{\smallskip}
    Spw. & Frequency & Flux density\\ 
    & (GHz) & (mJy)\\
    \noalign{\smallskip}
    \hline              
    \noalign{\smallskip}
    1 & 217.8 & $343.5\pm0.6$ \\
    2 & 220.0 & $345.1\pm0.9$ \\
    3 & 232.0 & $392.7\pm0.9$ \\
    4 & 234.5 & $403.8\pm0.9$ \\
    \noalign{\smallskip}
    \hline                  
    \end{tabular}
    \end{table}

    \begin{figure}[!htb]
        \centering
        \includegraphics[width=0.4\textwidth]{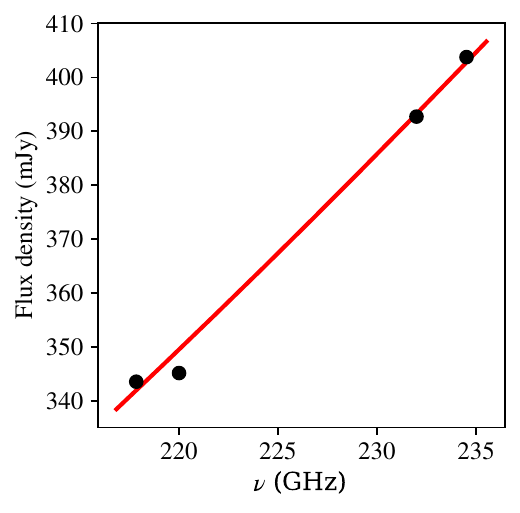}
        
        \caption{In-band flux density measurements for the four SPW continuum images (see Table~\ref{tab:spectral_index}). The red line represents the spectral index fit to the data ($\alpha=2.22\pm0.15$). Error bars are smaller than the marker size, and the error in the spectral index fit is dominated by the systematic relative flux uncertainty.}
        \label{fig:spectral_index}
    \end{figure}

\section{H30$\alpha$ versus $^{33}$SO$_2$ emission}

    In Sect. \ref{sect:h30a}, we discussed a possible contaminant in the H30$\alpha$ detection. As outlined by \citet{sanchez-monge2025} for the source W75N(B)-VLA3, this contaminant may be the line forest of $^{33}$SO$_2$. To test the validity of our H30$\alpha$ detection, we extracted the spectra at two cubes, one with no baselines removed, and another one with baselines shorter than 500~k$\lambda$ removed, to account for any possible molecular extended emission. In order to make a more robust comparison, we degraded the resolution to a circular beam of 80~mas, similar to the beam size of the cube including all baselines ($\theta_s=\mathrm{79~mas\times63~mas}$, P.A.=-80.9).
    
    Figure~\ref{fig:line_profile_33SO2} shows a comparison between the spectra extracted from the cube with no baseline removal (red lines) and the cube removing all baselines with projected lengths $<500$~k$\lambda$ (black lines). The top and bottom panels show the spectra extracted from a region of $0.5\arcsec\times0.5\arcsec$ and at the position of the peak of radio continuum emission, respectively.

    \begin{figure}[!htb]
        \centering
        \includegraphics[width=0.49\textwidth]{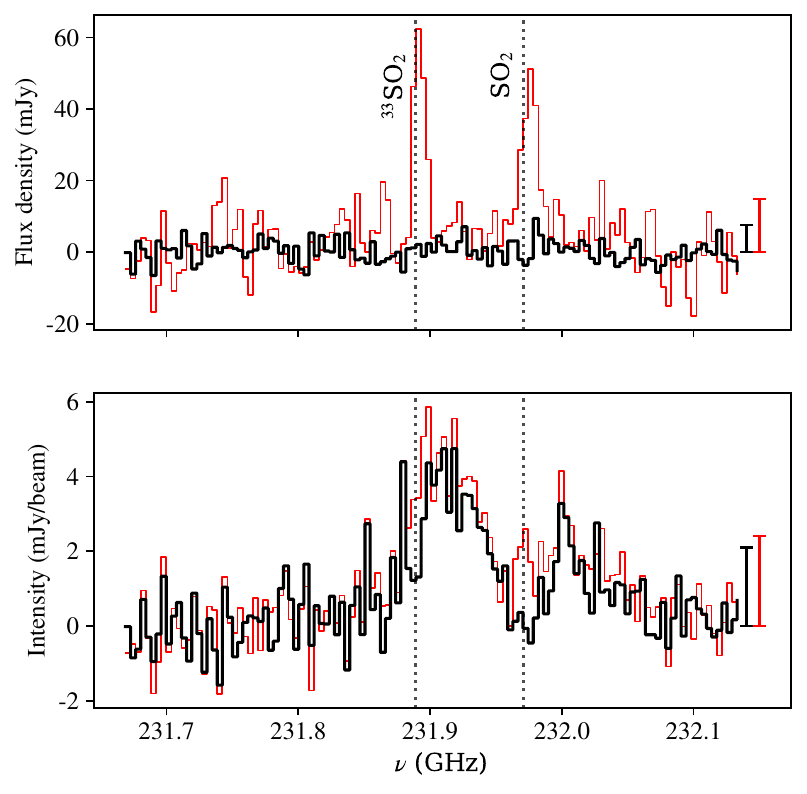}
        
        \caption{Spectra extracted at the continuum peak (see Table~\ref{tab:continuum_fit}) using the two cubes degraded to a common circular beam size of 80~mas. Red lines show the cube including all baselines, whereas black lines show the cube with baselines $<500~\mathrm{k\lambda}$ removed. \textit{Top:} Flux density integrated over a box of  $0.5\arcsec\times0.5\arcsec$. \textit{Bottom:} Intensity at the peak continuum pixel. The vertical dotted grey lines indicate the observed frequencies of the lines SO$_2$ ($14_{3,11}-14_{2,12};\, \nu_2=1$) (rest frequency $\sim$231.9805~GHz) and $^{33}$SO$_2$ ($3_\mathrm{9,K_c}-2_\mathrm{10,K_c}$) transitions (mean rest frequency $\sim$231.8984~GHz), considering the systemic velocity of the source \citep[$v_\mathrm{lsr}=+12.5~\mathrm{km~s^{-1}}$,][]{anez2020}. The vertical bars to the right of the spectra represent $3\sigma$ confidence level measured over line-free channels.}
        \label{fig:line_profile_33SO2}
    \end{figure}

\section{Moment 0 map of the secondary spectral feature}\label{sect:second_peak}

    In this Letter, we focus on the spectral feature centred at $v_\mathrm{LSR}\sim-14~\mathrm{km~s^{-1}}$, which we identify as the primary H30$\alpha$ component. However, as shown in Fig.~\ref{fig:line_profile}, a secondary peak is also clearly detected at $v_\mathrm{LSR}\sim-127~\mathrm{km~s^{-1}}$. To assess whether this feature traces an additional velocity component, we constructed a moment 0 (integrated intensity) map over the velocity range $-140 \leq v_\mathrm{LSR}~(\mathrm{km~s^{-1}}) \leq -115$. The resulting map is presented in Fig.~\ref{fig:mom0_secondpeak}. Although the emission is significantly weaker than that of the primary component, it exhibits a similar morphology, with a peak displaced towards the northeast and a marginal elongation in the same direction, consistent with the properties observed for the primary H30$\alpha$ component. This supports that this component also represents RRL emission, either from H30$\alpha$ or He30$\alpha$.

    \begin{figure}[!htb]
        \centering
        \includegraphics[width=0.35\textwidth]{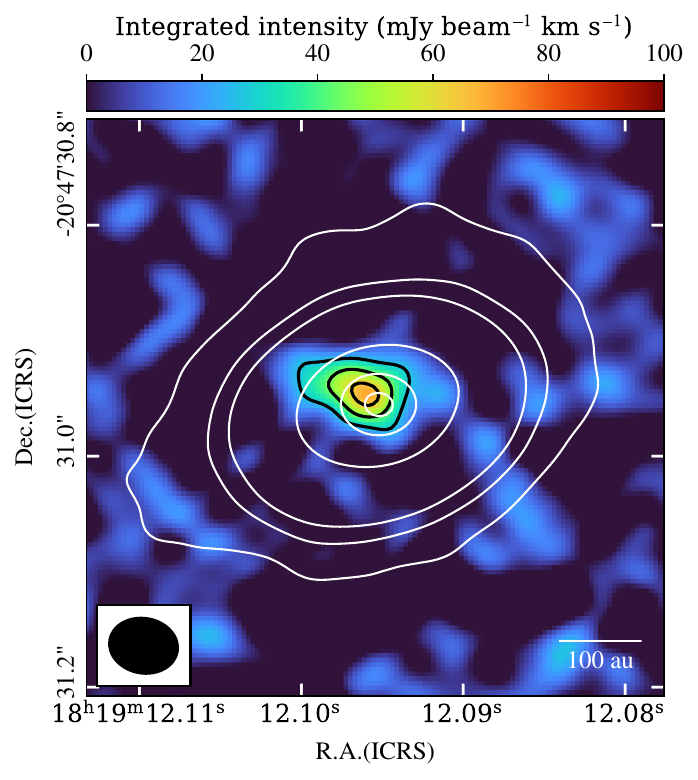}
        
        \caption{
        Moment 0 map for the secondary spectral feature integrated in the velocity range $-140\leq v_\mathrm{lsr}~(\mathrm{km~s^{-1}})\leq-115$ (see Fig.~\ref{fig:line_profile}). The thick black contours are $(3,5,7)\times\sigma_\mathrm{mom0}$ where $\sigma_\mathrm{mom0}=9.0~\mathrm{mJy~beam^{-1}~km~s^{-1}}$. The white contours are the same as in Fig.~\ref{fig:cont_mod_res}.}
        \label{fig:mom0_secondpeak}
    \end{figure}

\end{appendix}
\end{document}